\documentclass{emulateapj}

\usepackage{graphics}
\usepackage{epsfig}
\usepackage{natbib}
\shorttitle{CO \& H$_{2}$ Absorption in AA Tau}
\shortauthors{France et al.}
\begin{document}

\title{CO and H$_{2}$ Absorption in the AA Tauri Circumstellar Disk\altaffilmark{*}}


\author{
Kevin France\altaffilmark{1}, Eric B. Burgh\altaffilmark{1},
Gregory J. Herczeg\altaffilmark{2}, 
Eric Schindhelm\altaffilmark{1}, 
Hao Yang\altaffilmark{3},
Herv{\'e} Abgrall\altaffilmark{4},
Evelyne Roueff\altaffilmark{4},
Alexander Brown\altaffilmark{1}, 
Joanna M. Brown\altaffilmark{5},
Jeffrey L. Linsky\altaffilmark{3} 
}

\altaffiltext{*}{Based on observations made with the NASA/ESA $Hubble$~$Space$~$Telescope$, obtained from the data archive at the Space Telescope Science Institute. STScI is operated by the Association of Universities for Research in Astronomy, Inc. under NASA contract NAS 5-26555.}

\altaffiltext{1}{Center for Astrophysics and Space Astronomy, University of Colorado, 389 UCB, Boulder, CO 80309; kevin.france@colorado.edu}
\altaffiltext{2}{Max-Planck-Institut f\"{u}r extraterrestriche Physik, Postfach 1312, 85741 Garching, Germany}
\altaffiltext{3}{JILA, University of Colorado and NIST, 440 UCB, Boulder, CO 80309}
\altaffiltext{4}{LUTH and UMR 8102 du CNRS, Observatoire de Paris, Section de Meudon, Place J. Janssen, 92195 Meudon, France}
\altaffiltext{5}{Harvard-Smithsonian Center for Astrophysics, 60 Garden Street, MS-78, Cambridge, MA 02138, USA}


\begin{abstract}
The direct study of molecular gas in inner protoplanetary disks is 
complicated by uncertainties in the spatial distribution of the gas, the time-variability of the source, and the comparison of observations across a wide range of wavelengths.  Some of these challenges can be mitigated with far-ultraviolet spectroscopy.  Using new observations obtained with the {\it Hubble Space Telescope}-Cosmic Origins Spectrograph, we measure column densities and rovibrational temperatures for CO and H$_{2}$ observed on the line-of-sight through the AA Tauri circumstellar disk.  CO $A$~--~$X$ absorption bands are observed against the far-UV continuum.  The CO absorption is characterized by log$_{10}$($N$($^{12}$CO)) = 17.5~$\pm$~0.5 cm$^{-2}$ and $T_{rot}$(CO) = 500$^{+500}_{-200}$ K, although this rotational temperature may underestimate the local kinetic temperature of the CO-bearing gas.   We also detect $^{13}$CO in absorption with an isotopic ratio of~$\sim$~20.  We do not observe H$_{2}$ absorption against the continuum; however, hot H$_{2}$ ($v$~$>$~0) is detected in absorption against the Ly$\alpha$ emission line.  We measure the column densities in eight individual rovibrational states,  determining a total log$_{10}$(N(H$_{2}$)) = 17.9$^{+0.6}_{-0.3}$ cm$^{-2}$ with a thermal temperature of $T$(H$_{2}$) = 2500$^{+800}_{-700}$ K.  The high-temperature of the molecules, the relatively small H$_{2}$ column density, and the high-inclination of the AA Tauri disk suggest that the absorbing gas resides in an inner disk atmosphere.
If the H$_{2}$ and CO are co-spatial within a molecular layer $\sim$~0.6 AU thick, this region is characterized by $\langle$$n_{H_{2}}$$\rangle$~$\sim$~10$^{5}$ cm$^{-3}$ with an observed $\langle$CO/H$_{2}$$\rangle$ ratio of $\sim$~0.4.  
We also find evidence for a departure from a purely thermal H$_{2}$ distribution, suggesting that excitation by continuum photons and H$_{2}$ formation may be altering the level populations in the molecular gas.  
\end{abstract}

\clearpage

\section{Introduction}

Molecular hydrogen (H$_{2}$) is the primary constituent of giant planets, both in the solar system~\citep{stevenson82} and in extrasolar systems~\citep{sudarsky03}.  H$_{2}$ is typically assumed to make up the majority of the mass in disks 
where giant planets form.  Owing in part to its lack of permanent dipole moment, H$_{2}$ does not emit or absorb strongly in spectral regions easily accessible from the ground, and is therefore challenging to observe directly.  The second most abundant molecule in these disks, carbon monoxide (CO), is often observed as a surrogate for H$_{2}$~\citep{salyk07,rodriguez10}.  The use of CO as a tracer for H$_{2}$, and therefore as a tracer of the molecular gas mass of the inner disk, relies on a CO/H$_{2}$ conversion factor.  In practice, this conversion factor is only well-constrained observationally in diffuse and translucent interstellar clouds~\citep{burgh07}.  In dense clouds, model calculations of the CO/H$_{2}$ ratio (defined here as $N$(CO)/$N$(H$_{2}$)) tend towards 1~--~3~$\times$~10$^{-4}$ (see e.g., Visser et al. 2009 and references therein).  This is in agreement with an extrapolation of the translucent cloud trend, but has only been verified on one sightline~\citep{lacy94}.  The CO/H$_{2}$ ratio in low-mass protoplanetary disks, in particular at planet-forming radii ($a$~$<$~10 AU), is almost completely unconstrained by direct observation.\nocite{france11a}  

A better understanding of the relationship between H$_{2}$ and CO in the inner regions of protoplanetary disks can give a more complete picture of the composition and structure of planet-forming regions.  In addition to questions of abundance, the distribution of molecular gas and its physical characteristics are of interest for models of giant planet formation and their migration (see Najita et al. 2007 for a review of molecular gas in inner disks).\nocite{najita07}  The far-ultraviolet (far-UV) bandpass provides access to the strongest dipole-allowed bands systems of both H$_{2}$ and CO.  Due to large opacities of dust and gas, far-UV photons do not probe the disk midplane directly.  However, far-UV spectroscopy offers the most direct means of simultaneously measuring H$_{2}$ and CO in the upper layers of inner disks.

Far-UV spectroscopy has proven to be an effective technique for probing the 
environments of young circumstellar disks.  First 
with the {\it International Ultraviolet Explorer} ($IUE$)~\citep{brown81,valenti00}, then with the {\it Far-Ultraviolet Spectroscopic Explorer} ($FUSE$; Wilkinson et al. 2002; Herczeg et al. 2005)\nocite{wilkinson02,herczeg05} and the {\it Hubble Space Telescope}-Space Telescope Imaging Spectrograph ($HST$-STIS) and Goddard High-Resolution Spectrograph (GHRS;~Ardila et al. 2002; Herczeg et al. 2002; Walter et al. 2003), and today with $HST$-Cosmic Origins Spectrograph~(COS; Ingleby et al. 2011; France et al. 2011b).\nocite{ingleby11,france11a}  The majority of these observational efforts have focused on the detection of H$_{2}$ fluorescence from hot molecular, inner disk ($a$~$\lesssim$~2 AU; Herczeg et al. 2004) gas illuminated by Ly$\alpha$~\citep{ardila02} and other atomic lines~\citep{wilkinson02,france07b}.  The order of magnitude gains in spectroscopic sensitivity afforded by COS have made it possible to expand these studies to the ultraviolet emission lines of CO (France et al. 2011a), enabling direct comparison of the dominant molecular species in these objects.\nocite{herczeg04,france11b}  Recently, \citet{yang11} and \citet{france11b} have presented the first detections of H$_{2}$ and CO absorption, respectively, in the UV spectra of Classical T Tauri Stars (CTTSs).  In this paper, we present unique $HST$-COS observations of a sightline through the upper atmosphere of the AA Tauri circumstellar disk where we detect the absorption lines of $both$ H$_{2}$ and CO.  \nocite{ardila02,herczeg02,walter03} 

Our paper is laid out as follows: 
In \S2, we describe the COS observations and data reduction. In \S3, we discuss the qualitative features of the AA Tau spectrum and the data analysis performed to derive the H$_{2}$ and CO column densities in the AA Tau disk.  
We use these results to constrain the physical conditions of both molecules in \S4, including the spatial distribution of the gas and the role of UV photons in regulating the molecular level populations.  We present a brief summary of the paper in \S5. The details of our modeling and the non-detection of H$_{2}$ against the UV continuum of AA Tau are presented in an Appendix.

\section{$HST$-COS Observations of AA Tau} 

\subsection{AA Tau} 
AA Tau is a well-studied K7 classical T Tauri star in the Taurus Molecular Cloud with an age of 1--3 Myr~\citep{bouvier99,white01,kraus09}.  
The mass-accretion rate onto AA Tau based on $U$-band observations 
 has been observed to be in the range 
3.3~--~7.1~$\times$ 10$^{-9}$ $M_{\odot}$ yr$^{-1}$ (Valenti et al. 1993; Gullbring et al. 1998; White \& Ghez 2001). 
Both high-resolution dust disk imaging~\citep{andrews07} and 
scattered light models of the photopolarimetric variability of the circumstellar material  indicate a high disk inclination ($i$~$\approx$~75\arcdeg; O'Sullivan et al. 2005, see also Bouvier et al. 1999).\nocite{bouvier99,valenti93}   

Because of its high inclination, AA Tau has become the prototypical example of dipolar magnetospheric accretion onto the star (Bouvier et al. 2007; Donati et al. 2010).  Periodically the magnetospheric accretion column passes in our line of sight, producing redshifted absorption in Balmer lines.  This absorption often coincides with eclipses in the photometric light curve, which may be caused by a warp in the inner disk caused by the funnel flow.   The eclipse periodicity of 8.22 $\pm$ 0.03 days is likely similar to the stellar period.\nocite{bouvier07,donati10}

Due to its relatively high extinction ($A_{V}$~=~0.74; Gullbring et al. 1998), far-UV H$_{2}$ emission was not detected towards AA Tau by $IUE$~\citep{valenti00}. As predicted by these authors, this emission was below the sensitivity limit of $IUE$.  AA Tau exhibits a far-UV excess characteristic of young disks~\citep{ingleby09}. The higher-resolution UV spectra presented in this work reveal a wealth of molecular (H$_{2}$ and CO) and atomic emission features consistent with active accretion of a gas-rich disk (\S3.1).  \citet{najita09} present ground-based mid-IR observations of the molecular inner disk ($a$~$\lesssim$~0.5 AU), traced by CO fundamental ($\Delta$$v$~=~1) band emission and atomic emission ([\ion{Ne}{2}]) from somewhat larger radii.  Mid-IR spectra from the {\it Spitzer Space Telescope} show that the AA Tau disk has a rich chemistry of more complex molecules (H$_{2}$O and OH) and organics such as HCN, C$_{2}$H$_{2}$ and CO$_{2}$~\citep{carr08}.  \citet{bethell09} use these results to demonstrate that water vapor formation may be common in CTTS disks and that water may be the dominant reservoir for oxygen in these systems.  As the most abundant and stable molecules in the inner disk, H$_{2}$ and CO are the building blocks for this complex chemistry. 

\begin{figure}
\begin{center}
\epsfig{figure=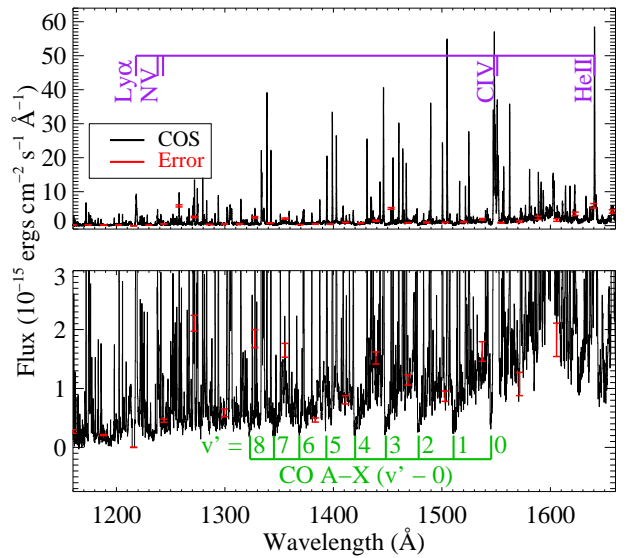,width=2.75in,angle=90}
\vspace{+0.15in}
\caption{
\label{cosovly} The COS far-UV spectrum of AA Tauri. The data are binned to one spectral resolution element for display and representative error bars are shown in red.  In the top panel, we show the full dynamic range spectrum, with the strongest atomic emission lines labeled in purple.  All other strong emission features in the 1160~--~1660~\AA\ bandpass are attributable to H$_{2}$ fluorescence.  The bottom panel shows the weak emission and absorption features as well as the far-UV continuum.  The majority of the weak emission lines are fluorescently excited CO and H$_{2}$.  The detected CO $A$~--~$X$ absorption bands are labeled in green.    
 }
\end{center}
\end{figure}

\subsection{Observations}
AA Tau was observed by COS on 2011 January 06 and 07 (program ID 11616) during four contiguous spacecraft orbits.  Target acquisition was performed in the MIRRORB near-UV imaging mode through the primary science aperture. The far-UV spectrum was obtained using multiple settings of the G130M and G160M gratings in order to achieve continuous, moderate resolution ($\Delta$$v$~$\approx$~17 km s$^{-1}$)\footnote{The COS LSF experiences a wavelength dependent non-Gaussianity due to the introduction of mid-frequency wave-front errors produced by the polishing errors on the $HST$ primary and secondary mirrors; {\tt http://www.stsci.edu/hst/cos/documents/isrs/}.} spectra in the 1133~$\leq$~$\lambda$~$\leq$~1795~\AA\ bandpass.  Two central wavelength settings were used in the G130M mode ($\lambda$1291 and  $\lambda$1327) at two focal plane offset positions (FP-POS = 1,4).  The G160M spectra were acquired at three central wavelengths ($\lambda$1577, $\lambda$1600, $\lambda$1623) at the default focal plane position (FP-POS = 3).  In addition to providing continuous wavelength coverage across the far-UV bandpass, the use of multiple grating settings minimizes the impact from fixed pattern noise inherent to the microchannel plate detector. A complete description of the COS instrument and on-orbit performance characteristics can be found in~\citet{osterman11}.
The one-dimensional spectra produced by the COS calibration pipeline, CALCOS, were aligned and coadded using the custom software procedure described by~\citet{danforth10}.

\section{Results and Analysis}

\subsection{AA Tau Emission Spectrum}

AA Tau displays a rich emission line spectrum across the far-UV bandpass (see Figure 1).   The majority of the lines can be attributed 
to H$_{2}$ fluorescence, excited by shock-driven \ion{H}{1} Ly$\alpha$ (see e.g., Herczeg et al. 2002).  The signature of photo-excited
CO $A$~--~$X$ emission~\citep{france11b} is also present.  
AA Tau displays a prominent 1550~--~1650~\AA\ excess that is attributed to a combination of collisionally excited H$_{2}$ and CO $A$~--~$X$ (0~--~1) band emission that may be  fluorescently pumped through a coincidence with \ion{C}{4}~\citep{bergin04,france11a,france11b}.  

Species tracing high-formation temperature regions ($T$~$>$~8~$\times$~10$^{4}$ K) are present in the spectrum, including \ion{N}{5} $\lambda$1239, 1243, \ion{C}{4} $\lambda$1548, 1550, and \ion{He}{2} $\lambda$1640~\AA.  
These lines are likely formed in the funnel flows~\citep{muzerolle01} and active chromosphere/transition region~\citep{krull00,bouvier07}.
Lower excitation species such as \ion{Si}{3} $\lambda$1206, \ion{C}{2} $\lambda$1334, 1335, and \ion{Si}{2} $\lambda$1526, 1533~\AA\ do not appear to be strong in the spectrum of AA Tau.  The atomic species and CO fluorescence merit further investigation, however in this paper we focus on the molecular absorption spectrum through the AA Tau disk.   

In the following subsections, we describe the analysis of the H$_{2}$ and CO absorption features, recently identified for the first time in CTTS spectra~\citep{yang11,france11b}.   We observe H$_{2}$ absorption lines superimposed upon the shock-created Ly$\alpha$ profile, and we describe the reconstruction of this profile and column density analysis of the absorbing gas in \S3.2.  We also detect CO $A$~--~$X$ absorption against the far-UV continuum from AA Tau, and in \S3.3, we describe the column density determination for this gas.

\subsection{H$_{2}$ Absorption Against the Ly$\alpha$ Profile in AA Tau}

H$_{2}$ absorption has been observed against both the stellar continuum and transition region emission lines in Herbig Ae/Be stars~\citep{roberge01,zaidi08}. Molecular absorption in  UV spectra of high-inclination CTTSs was beyond the detection limit of space-borne instruments prior to the installation of COS.  \citet{herczeg04} presented a detailed analysis of H$_{2}$ in the disk of TW Hya, including a reconstruction of the Ly$\alpha$ line profile and determination of the H$_{2}$ opacity in the disk atmosphere.  Their work suggested that H$_{2}$ absorption could in principle be detected against the Ly$\alpha$ emission line in gas-rich disks.  No other disks were observed by STIS at high enough S/N and resolving power to detect these absorptions, however, H$_{2}$ absorption in CTTS systems has recently been detected against the wings of the strong Ly$\alpha$ emission line~\citep{yang11}.  H$_{2}$ absorption is observed against the red-wing of the Ly$\alpha$ profile in AA Tau, shown in Figure 2. We do not observe H$_{2}$ absorption against the far-UV continuum of AA Tau, and upper limits on the H$_{2}$ for this absorption can be found in the Appendix.  In the following subsections, we describe their measurement and derive the column density of hot H$_{2}$ in the AA Tau disk.  
 
\begin{figure*}
\begin{center}
\epsfig{figure=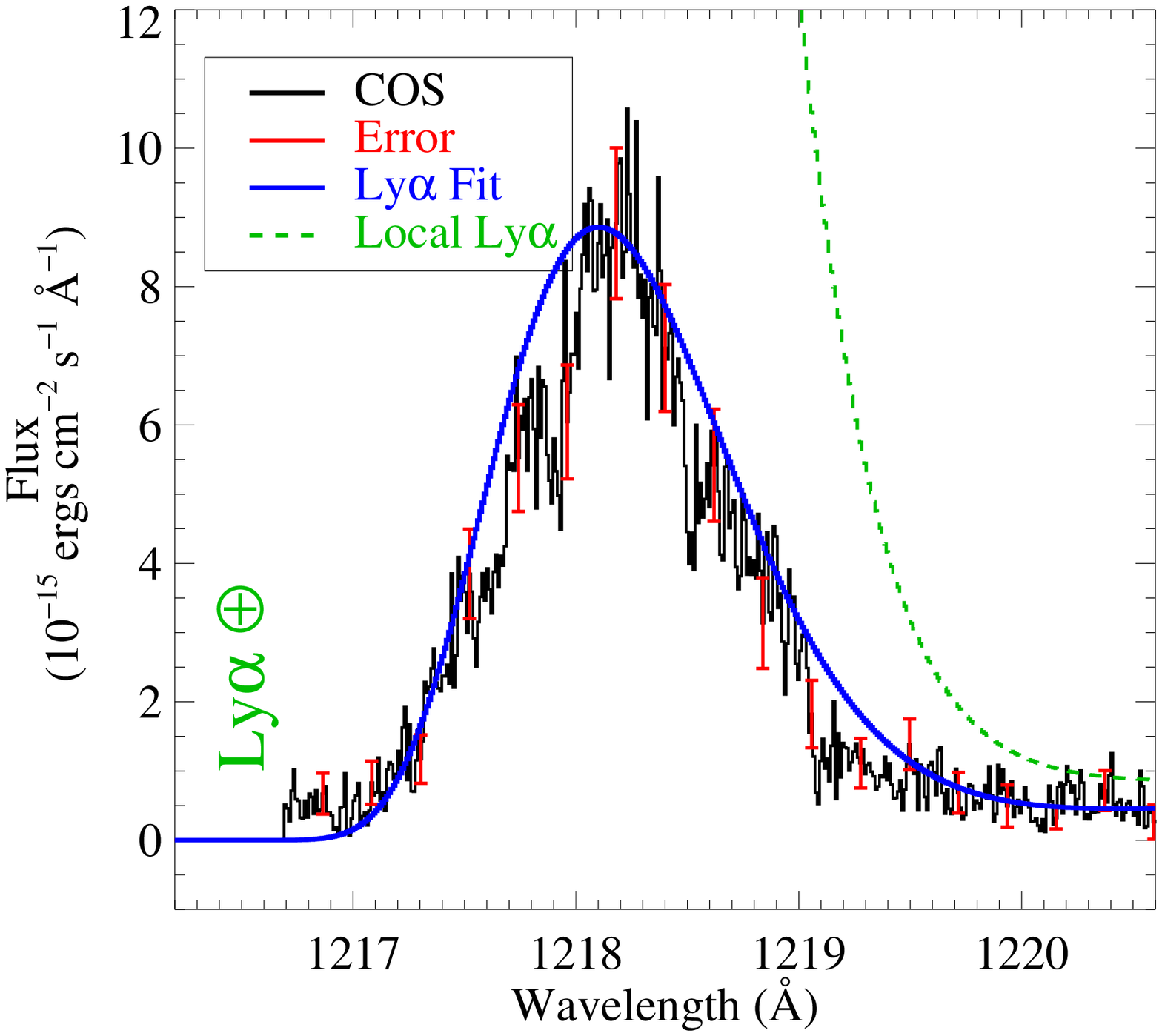,width=2.4in,angle=90}
\epsfig{figure=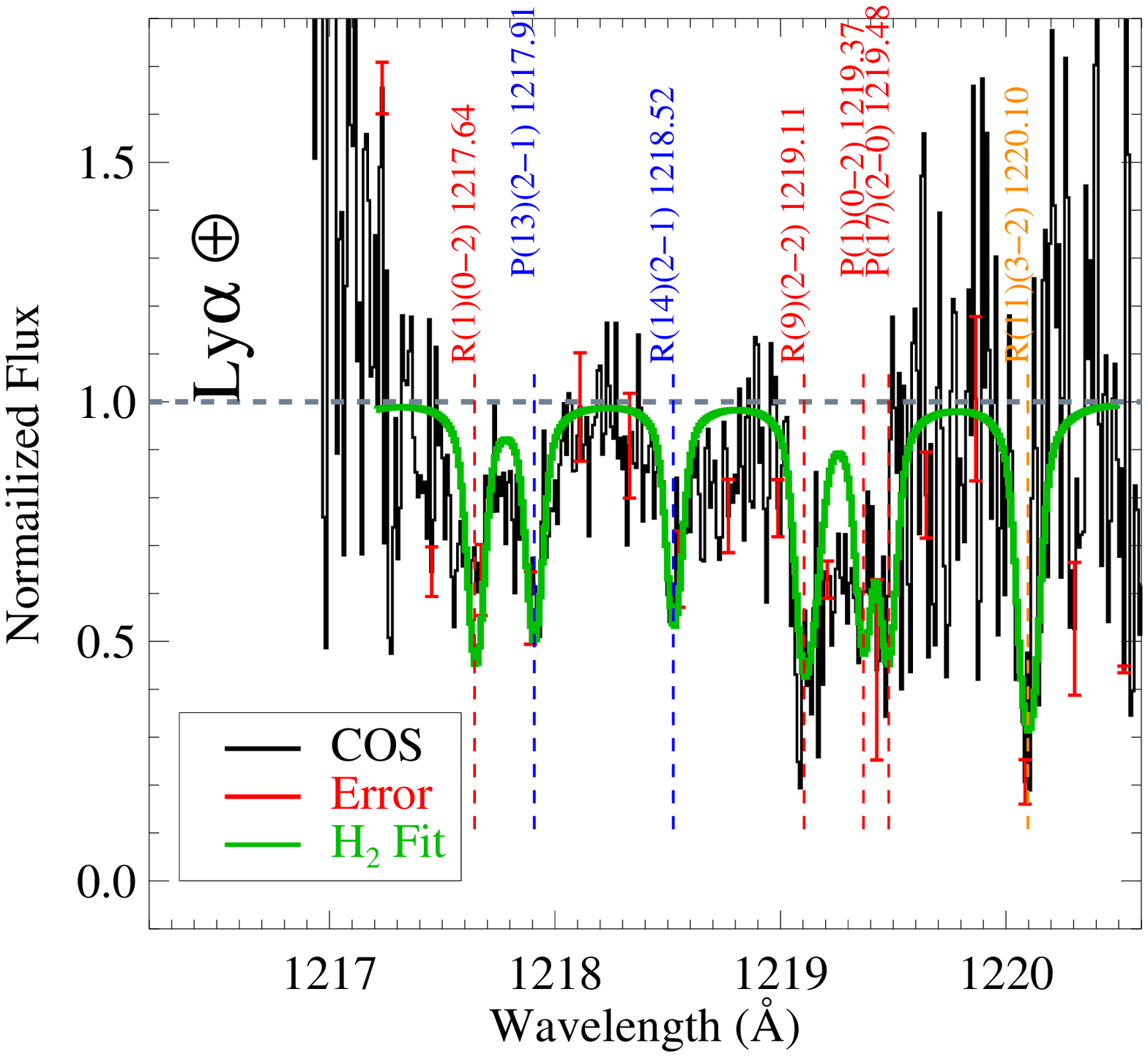,width=2.4in,angle=90}
\vspace{+0.25in}
\caption{
\label{cosovly} (a~--~$left$) The red wing of the stellar + shock Ly$\alpha$ profile of AA Tau. The data are displayed in black with representative error bars plotted in red.  The Ly$\alpha$ model, including a fit to the interstellar \ion{H}{1} column density, is shown in blue.  The dashed green line represents the local Ly$\alpha$ radiation field incident on the disk.  (b~--~$right$) The H$_{2}$ absorption spectrum after dividing out the Ly$\alpha$ model.  The data are again shown in black with representative error bars shown in red.  The best fit H$_{2}$ absorption spectrum for $b_{H2}$~=~5 km s$^{-1}$ is shown in green.   Prominent absorption features are labeled on the figure.  Table 1 lists the best fit H$_{2}$ column densities for all observed lines.  
 }
\end{center}
\end{figure*}

\subsubsection{Ly$\alpha$ Profile Reconstruction}

The H$_{2}$ transmission spectrum is the ratio of an incident Ly$\alpha$ profile to the data.   We created a simple empirical fit to the Ly$\alpha$ emission line + interstellar absorption in order to reproduce the red side of the line profile.  First, we remove the 1214.5~--~1216.8~\AA\ flux, which is dominated by geocoronal \ion{H}{1} emission.  
H$\alpha$ and H$\beta$ spectra of AA Tau display absorption characteristic of a an inner disk wind~\citep{bouvier07}, and the blue side of the Ly$\alpha$ line profile is heavily suppressed; its flux is a factor of $\sim$~10~lower than the red wing of the line.  This is presumably due to either an accelerating low-density outflow or an optically thick but patchy outflow at a single velocity.  For the present work, we assume that the majority of the 
Ly$\alpha$ emission profile is formed near the star and create a baseline on which to measure the H$_{2}$ absorption spectrum.  

We found that a single Gaussian emission line combined with a rest-velocity \ion{H}{1} absorption component could provide satisfactory fits to the observed Ly$\alpha$ spectrum, shown in Figure 2a.  The Gaussian emission line is characterized by a peak amplitude of 2.65~$\times$~10$^{-12}$ erg cm$^{-2}$ s$^{-1}$ \AA$^{-1}$, a FWHM of 620 km s$^{-1}$, and a continuum level of 9.5~$\times$~10$^{-16}$ erg cm$^{-2}$ s$^{-1}$ \AA$^{-1}$.  
An interstellar \ion{H}{1} profile characterized by $N$(HI) = 4.3~$\times$~10$^{20}$ cm$^{-2}$ and $b_{HI}$ = 10 km s$^{-1}$ provides a good fit to the data.  It is interesting to note that this column density corresponds to an extinction of $A_{V}$~=~0.28, assuming $R_{V}$~=~3.1 and the interstellar gas-to-dust relationship between $N$(\ion{H}{1}) and $E(B~-~V)$~\citep{bohlin78}.
That is a factor of $\sim$~3 lower than the $A_{V}$ typically assumed for the AA Tau sightline, suggesting that a large fraction of the extinction derived from optical and near-IR studies arises in circumstellar material with a gas/dust ratio less than that in the general ISM.   
The parameters for the emission line may be an oversimplification of actual Ly$\alpha$ profile in AA Tau, but the interstellar values are constrained by the inner and outer edges of the observed \ion{H}{1} profile and are therefore fairly robust.  There was a minor residual shape to the fit/data H$_{2}$ transmission spectrum.  This $\approx$~10~\% slope was fit by a spline function and removed.   

\subsubsection{$N$(H$_{2}$,$v$,$J$)} 

The AA Tau H$_{2}$ absorption spectrum is displayed in Figure 2b. 
The strongest H$_{2}$ absorption lines are labeled.
After dividing out the model Ly$\alpha$ emission line, there is~2.8~\AA\ of spectral region with high enough S/N for H$_{2}$ profile fitting.     We created a multi-component H$_{2}$ fitting routine to measure the column density in the nine absorption lines into the Lyman 
($B$$^{1}\Sigma^{+}_{u}$~--~$X$$^{1}\Sigma^{+}_{g}$) band system 
 that contribute significantly to the molecular opacity between 1217.48 and 1220.30~\AA\footnote{We note that there is a possible absorption on the blue side of Ly$\alpha$ from the $B$~--~$X$ (1~--~1) P(11) line at $\lambda_{obs}$~=~1212.55~\AA, but the absorption depth is at the RMS noise level of the data at these wavelengths.}.  We create the intrinsic line profiles as a function of column density in a given rovibrational [$v$,$J$] level.  These functions are added in optical depth space, a transmission curve is created, and this function is then convolved with the most recent COS line-spread function~\citep{kriss11} for comparison with the absorption spectrum.  The best-fit multi-line H$_{2}$ model is found using the MPFIT routine for a fixed $b_{H2}$~=~5 km s$^{-1}$ (see the Appendix for details of the H$_{2}$ model).  Initial conditions were determined by first manually fitting the H$_{2}$ spectrum, and in order to remove bias introduced by these choices, a grid of initial conditions were searched.  This method produced a total of $>$ 6000 individual realizations of the spectral fitting, and the best-fit column densities ($N$(H$_{2}$,$v$,$J$)) were taken from the mean of these distributions.  

\begin{deluxetable}{lccccc}
\tabletypesize{\normalsize}
\tablecaption{H$_{2}$ Absorption Lines observed against Ly$\alpha$. \label{lya_lines}}
\tablewidth{0pt}
\tablehead{
\colhead{Line ID\tablenotemark{a}} & \colhead{$\lambda_{rest}$} & 
\colhead{log $N$(H$_{2}$,$v$,$J$)\tablenotemark{b}} & 
\colhead{$f$}  &  \colhead{$EW$}  & \colhead{$E$\tablenotemark{c}} \\ 
& (\AA) & (cm$^{-2}$)  &  &  (m\AA) & (eV)}
\startdata
(0~--~2)R(1) & 1217.64 & 14.91$^{+0.12}_{-0.32}$  & 0.0289 & 59$^{+4}_{-11}$ &  1.02 \\
(2~--~1)P(13) & 1217.91 & 14.86$^{+0.20}_{-0.42}$ & 0.0192 & 52$^{+7}_{-19}$ &  1.64  \\
(2~--~1)R(14) & 1218.52 & 14.81$^{+0.18}_{-0.28}$ & 0.0181 & 49$^{+7}_{-13}$ &  1.79  \\
(0~--~2)R(2) & 1219.09 & 14.22$^{+0.60}_{-0.28}$  & 0.0255 & 27$^{+28}_{-11}$ &  1.04 \\
(2~--~2)R(9) & 1219.10 & 14.35$^{+0.60}_{-0.28}$  & 0.0318 & 40$^{+22}_{-17}$ &  1.56   \\
(2~--~2)P(8) & 1219.16 & 14.13$^{+0.60}_{-0.28}$  & 0.0214 & 8$^{+40}_{-3}$ &  1.46  \\

(0~--~2)P(1) & 1219.37 & 14.91$^{+0.12}_{-0.32}$  & 0.0149 & 50$^{+4}_{-11}$ & 1.02 \\

(2~--~0)P(17) & 1219.48 & 15.60$^{+0.50}_{-0.30}$ & 0.0040 & 54$^{+23}_{-22}$ &  1.85  \\
(3~--~2)R(11) & 1220.11 & 16.34$^{+0.78}_{-0.78}$ & 0.0213 & 92$^{+15}_{-18}$ &  1.80 
\enddata
\tablenotetext{a}{Transitions are for the $B$$^{1}\Sigma^{+}_{u}$~--~$X$$^{1}\Sigma^{+}_{g}$ H$_{2}$ band system. } 
 \tablenotetext{b}{Column density and equivalent width calculations assume $b_{H2}$~=~5 km s$^{-1}$, see \S3.2.2. } 
 \tablenotetext{c}{Column densities, oscillator strengths, equivalent widths, and energy levels refer to the rovibrational level of the ground electronic state out of which the Ly$\alpha$ photons are absorbed.} 
\end{deluxetable}

This multi-component fitting method was mostly insensitive to the exact initial conditions, except for the (0~--~2)R(2) and (2~--~2)R(9) levels, whose wavelengths differ by only 0.01~\AA\ ($\lambda_{rest}$~=~1219.09 and 1219.10~\AA, respectively).  The total column density at this wavelength was robust, however the relative columns in the two lines was not.  We weighted the individual columns by the product of the oscillator strength and the relative populations of the two levels at a fiducial hot H$_{2}$ temperature ($T$(H$_{2}$) = 2500 K).  The oscillator strengths and populations of the two lines are [$f_{R(2)}$ = 25.5~$\times$~10$^{-3}$, $P_{R(2)}$ = 5.76~$\times$~10$^{-4}$] and [$f_{R(9)}$ = 31.8~$\times$~10$^{-3}$, $P_{R(9)}$ = 6.24~$\times$~10$^{-4}$], respectively.  $N$(2,2) contributes 0.425 and $N$(2,9) contributes 0.575 to the total 1219.10~\AA\ column.  The (2~--~0)P(17) line is located at the inflection point between the Ly$\alpha$ line and the far-UV continuum, complicating its line-shape.  This line was fit manually ($N$(0,17) = 4~$\times$~10$^{15}$ cm$^{-2}$) and this value was fixed during the fitting procedure to reduce run time.   The best-fit H$_{2}$ column densities are presented in Table 1.  

The error bars presented in Table 1 were determined from the minimum and maximum deviations from the mean column densities of the best-fit distribution.  In cases where the fitting uncertainty errors were less than the 1-$\sigma$ error bars on the data (shown in Figure 2b), we took the errors to be the range of column densities that could be accommodated within the uncertainty on the data.  This approach gives a conservative estimate of the uncertainties on the measured column density.  Significantly larger $b$-values would imply turbulent velocities greater than the isothermal sound speed for $T$~$\lesssim$~4000 K, however, in order to fully understand the parameter space, we also determined the best-fit parameters assuming $b_{H2}$~=~10 km s$^{-1}$.  As expected, increasing $b$ reduces the best-fit column densities.  We found that the column densities were typically 0.1~--~0.3 dex smaller at high $b$.  Two exceptions to this trend were the columns of the (2~--~2)P(8) (which increased by $\approx$~0.6 dex) and (3~--~2)R(11) lines (which decreased by $\approx$~1.3 dex).  The equivalent widths were calculated from the best-fit column densities, assuming that the observed H$_{2}$ lines are located on the ``flat part'' of the curve-of-growth, and are displayed in Table 1. 

\subsubsection{Thermal H$_{2}$ Population}

The measured column densities, normalized by the degeneracy of the state ($g_{J}$), are plotted as a function of excitation temperature ($T_{exc}$~=~$E(v,J)$/$k_{B}$, where $k_{B}$ is the Boltzmann constant) in Figure 3.  We label the absorbing levels in the ground electronic state as [$v$,$J$] and the rovibrational levels of the $B$$^{1}\Sigma^{+}_{u}$ electronic state as [$v^{'}$,$J^{'}$]\footnote{The rovibrational levels of the ground electronic state following fluorescence would then be [$v^{''}$,$J^{''}$].}.
The energy levels for the individual rovibrational levels were taken from~\citet{dabrowski84}.  Gas in thermal equilibrium is well-described by a single Boltzmann distribution, which would appear as a straight line in Figure 3, while deviations from this thermal distribution indicate that non-thermal processes are partially responsible for the level populations in the H$_{2}$ gas.   Non-equilibrium photo-excitation and excess energy associated with H$_{2}$ formation on grain surfaces are generally assumed to be the dominant non-thermal H$_{2}$ excitation processes in interstellar clouds~\citep{spitzer73} and protoplanetary disks~\citep{nomura05}, and it seems plausible that both are at work in the molecular disk of AA Tau (see \S4.1). 

The H$_{2}$ population is roughly consistent with a thermal distribution for $T_{exc}$~$<$~2~$\times$~10$^{4}$ K.  Assuming that the lowest energy states observed in absorption represent a thermal H$_{2}$ population, these lines are used to determine the total column density of this gas.  Figure 3 shows three thermal distribution curves that correspond to the best-fit, upper and lower limits to the thermal H$_{2}$ (the solid pink, dashed blue, and dashed green lines), normalized to the measured column in the [$v$,$J$] = [2,1] state.  
It may be that the [2,1] state itself is not thermally populated, but in the absence of lower-excitation constraints (most of the lines that would constrain the $T_{exc}$~$<$~5000 K portion of the excitation diagram are located at wavelengths shorter than the COS G130M band), we choose the lowest observed energy level for the normalization.  
With this caveat about the normalization in mind, we find that the thermal population can be characterized by $T$(H$_{2}$)~=~2500$^{+800}_{-700}$ K.  These values are consistent with the thermal origin of the Ly$\alpha$-pumped H$_{2}$ emission proposed by~\citet{herczeg04}.  The total column density of the thermal H$_{2}$ is thus log$_{10}$($N$(H$_{2}$)) = 17.86$^{+0.62}_{-0.31}$ cm$^{-2}$.  
We will present a discussion of the possible evidence for non-thermal H$_{2}$ level populations in \S4.1. 

\vspace{+0.5in}

\subsection{CO Absorption}

While H$_{2}$ absorptions from the ground vibrational state
($v^{'}$~--~0) are located at $\lambda$~$\lesssim$~1110~\AA, the
$A$~--~$X$ absorption system of CO spans the COS far-UV bandpass,
extending blueward from the (0~--~0) band at
$\lambda$~$\sim$~1544~\AA.  We identify nine CO bands from $v^{'}$ =
0~--~8 in the COS spectra of AA Tau (see Figure 1).  Spectral blending
with photo-excited H$_{2}$ as well as emission from \ion{Si}{4}
$\lambda$1394 and \ion{C}{4} $\lambda$1548 limit the number of bands
that can be used for spectral fitting\footnote{\citet{france11b} suggested that \ion{C}{4}
may be a photo-excitation source for CO in these systems, and that possibility 
will be explored in a future work.}, and   
the low flux levels
($F_{\lambda}$~$\lesssim$~1~$\times$~10$^{-15}$ erg cm$^{-2}$ s$^{-1}$
\AA$^{-1}$) also make detailed modeling challenging.  Broad CO $A$~--~$X$
absorption lines were first identified in the UV spectrum of HN
Tau~\citep{france11b}, and we employ a similar technique for fitting
the CO absorption spectrum of AA Tau.  To date, all CTTS systems that
show $A$~--~$X$ absorption have very broad CO profiles due to blending of closely-spaced rotational lines (suggesting CO temperatures of several hundred K) and relatively high disk inclinations, arguing that the molecules are located in a warm inner
disk.

Each CO band was extracted from the spectrum and a low-order
polynomial was fit to the most uncontaminated nearby portions of the
spectrum for continuum normalization.  Individual band velocities
relative to rest were determined to correct for any relative offsets
introduced by wavelength calibration errors or in the co-addition
process.  Profile fitting was used to determine the relevant
parameters: N($^{12}$CO), N($^{13}$CO), T$_{rot}$, and Doppler
$b$-value.  Because the thermal width of CO is $\approx$~4 times smaller
than that of H$_2$, we restricted our fitting to $b_{CO}$~$\leq$~2 km
s$^{-1}$.

CO absorption band profiles were modeled using ro-vibrational line
wavelengths and oscillator strengths from Eidelsberg (private
communication) for $^{12}$CO.  For $^{13}$CO, wavelengths were
determined from the ground state energies calculated from the
mass-independent Dunham coefficients of George et al. (1994) and the
A~$^1\Pi$ energy levels of Haridass \& Huber (1994), while the
oscillator strengths used were from Eidelsberg et al. (1999).  The
wavelengths and oscillator strengths for the perturbations, which show
up more prominently at higher column densities, were taken from
Eidelsberg \& Rostas (2003).  In the fitting process, we assumed the
ground states for both $^{12}$CO and $^{13}$CO were populated
following a single rotational temperature, T$_{rot}$(CO).

Of all the bands, the (1-0), (2-0), (3-0), (4-0), (7-0) and (8-0) were
the least contaminated by other spectral features so we compared
these bands simultaneously with a grid of absorption line models.
Best-fit parameters were determined using a chi-squared metric,
limiting the fitting to normalized flux values below 1.1; this
minimized the diluting effect the strong emission features would have
on the chi-squared value.  Our analysis finds that the CO absorption
is characterized by log$_{10}$($N$($^{12}$CO)) = 17.5~$\pm$~0.5
cm$^{-2}$, log$_{10}$($N$($^{13}$CO)) = 16.2~$\pm$~1.0 cm$^{-2}$,
$b_{CO}$~=~1.2~$\pm$~0.4 km s$^{-1}$, and $T_{rot}$(CO) = 500$^{+500}_{-200}$
K.  The best-fit CO model for several bands is shown overplotted in
orange on the COS spectrum in Figure 4.

The large errors in our best-fit parameters are indicative of the
difficulties inherent in fitting the CO absorption in data such as
these.  Several systematic effects contribute to the error.  Due to
the large number of emission features, finding suitable uncontaminated
spectral regions for continuum normalization was difficult.  
The (2~--~0) band shown in Figure 4 ($upper$~$right$) provides an example of 
the challenges associated with continuum placement in CTTSs.  
These emission features, coupled with the low signal-to-noise, also enhance
the error in our temperature fit; as rotational excitation temperature
increases, the profile near the low-lying J levels, where the lines
may be experiencing more saturation effects, changes less than at
higher-lying J levels, which lie at longer wavelengths.  The result is
a larger error bar to higher temperatures.  At the lower ($v^{'}$~--~0) bands, the
$^{13}$CO tends to overlap with the $^{12}$CO and as such tends to
complicate the fit; however, for the (7-0) and (8-0) bands, the
$^{13}$CO is cleanly separated and detected.
\nocite{eidelsberg99,eidelsberg03,haridass94,george94}

\begin{figure}
\begin{center}
\epsfig{figure=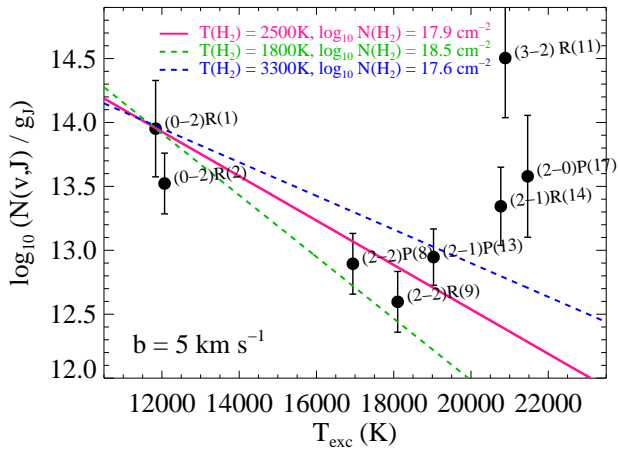,width=2.65in,angle=90}
\vspace{+0.15in}
\caption{H$_{2}$ excitation diagram based on measured rovibrational state column densities (Table 1).   
All absorption lines are for transitions to the Lyman band ($B$~--~$X$). 
Several high-$J$ levels at $T_{exc}$~$>$~2~$\times$~10$^{4}$ K show evidence for non-thermal excitation.   Thermal population curves (normalized to the column density in the [$v$~=~2, $J$~=~1] level) for rovibrational excitation temperatures corresponding to the best-fit, minimum, and maximum values are shown in pink, green, and blue respectively.   These curves imply total thermalized column densities of $N$(H$_{2}$) = 7.2, 30, and 3.5~$\times$~10$^{17}$ cm$^{-2}$ for the three curves.  
 }
\end{center}
\end{figure}

\subsubsection{Isotopic Fractionation}  

That $^{13}$CO absorption is required to fit the observed
absorption profiles is not unexpected considering the high $^{12}$CO
column densities observed in AA Tau~\citep{sonnentrucker07}.  The
best-fit log$_{10}$($N$($^{13}$CO))~=~16.2~$\pm$~1.0~cm$^{-2}$,
implies a $^{12}$CO/$^{13}$CO ratio of ~$\sim$~20, though we caution
that the uncertainties on both quantities are large ($^{12}$CO/$^{13}$CO~$\sim$~20$^{+600}_{-19}$) .  This ratio is
smaller than the local interstellar $^{12}$CO/$^{13}$CO
of~$\approx$~70 (Sheffer et al. 2007; see also the review by Wilson \&
Rood 1994) and the isotopic ratios derived from IR observations of young stellar objects ($^{12}$CO/$^{13}$CO~$\approx$~100; Smith et al. 2009), although the discrepancies are not statistically significant.\nocite{sheffer07,wilson94,smith09} We briefly consider the primary
processes that are expected to influence the isotopic fraction in
molecular environments.

Isotopic charge exchange enhances the population of
$^{13}$CO~\citep{vand88}, becoming efficient at kinetic
temperatures~$\lesssim$~35 K.  While the CO band widths strongly rule
out $T_{rot}$(CO)~$<$~100 K, turbulent mixing in the disk (e.g., Ilger
et al. 2004) could bring $^{13}$CO formed closer to the midplane to
the upper disk where we observe the absorbing gas.\nocite{ilger04}
$^{13}$CO is also selectively photodissociated by UV photons at the
surface of molecular clouds~\citep{bally82} due to the faster onset of
$^{12}$CO self-shielding, and this process enhances the relative population of $^{12}$CO  in these environments. 
While the far-UV radiation field at the
surface of a CTTS disk is intense~\citep{fogel11,france11a}, the
absorption bands with the largest $^{13}$CO dissociation probabilities
are located at $\lambda$~$<$~1060~\AA~\citep{vand88}, where the continuum is weak.  

A $^{12}$CO/$^{13}$CO
ratio of $\sim$~20 is marginally consistent with lowest inner disk carbon fractions
computed by~\citet{woods09}.  Their models suggest that our
observed $^{12}$CO/$^{13}$CO ratios would be expected closer to the
disk midplane (for the reasons noted above).  Therefore, if low
fractional ratios are found in the warmer upper disk, then this
material was likely formed in cooler, shielded regions and has
subsequently been vertically transported through the disk.  
It is also intriguing to note that the models presented by~\citet{visser09}
predict isotopic fractions $\lesssim$~30 for $N$(H$_{2}$)~$\gtrsim$~10$^{21}$ cm$^{-2}$.  
Assuming a typical interstellar CO/H$_{2}$ ratio (10$^{-4}$), $N$(H$_{2}$)~$\gtrsim$~10$^{21}$ cm$^{-2}$ is approximately the expected H$_{2}$ column associated with our
measured value of $N$(CO) = 10$^{17.5}$ cm$^{-2}$ (see \S4.3).


\section{Discussion}

\subsection{Non-Thermal H$_{2}$ Population}  

These data most likely probe a sightline through the upper atmosphere of the AA Tauri disk, a region where non-equilibrium processes are expected to be important.  \citet{france11b} suggested that non-thermal populations could be relevant for both the CO and H$_{2}$ populations in inner CTTS disks (see also the protoplanetary disk H$_{2}$ models of Nomura \& Millar 2005).  Their work was based primarily on model-dependent emission line analyses and analogy to the better-studied case of molecular excitation in photodissociation regions.  \citet{ardila02}, using fluorescently excited H$_{2}$ emission lines to reconstruct the ground state level populations in a sample of five CTTSs, found that the H$_{2}$ was considerably out of thermal equilibrium.   However, more recent analyses based on $HST$-STIS observations of CTTSs suggest that these levels are consistent with a thermal population~\citep{herczeg04,herczeg06}.  
 It should be emphasized that all of the studies cited above deal with molecular $emission$ lines, and it is not yet clear whether the CO and H$_{2}$ emission are tracing the same parcels of gas as the CO and H$_{2}$ absorption.  

Using a direct measurement of H$_{2}$ and CO absorption lines along the same line-of-sight and in the same observation, we have a much more straightforward observational basis for understanding the molecular processes operating in protoplanetary disk atmospheres.  The deviations from the thermal distributions in Figure 3 could be direct evidence for non-thermal excitation in the molecular disk atmosphere.  It is premature to draw broad conclusions until this behavior is observed in similar systems because we only detect three states with strongly non-thermal behavior, and two of these (2~--~0)P(17) and (3~--~2)R(11) are measured in regions of low S/N.  However, the excitation diagram presented in Figure 3 is intriguing, and we briefly describe possible mechanisms for this excitation in the following paragraphs. It is also interesting to note that we only measure significant deviations from a thermal population at excitation temperatures $T_{exc}$~$>$~2~$\times$~10$^{4}$ K, approximately the same energy cut-off where~\citet{ardila02} saw the largest non-thermal excursions.  

There are two main mechanisms that can drive the H$_{2}$ level populations out of thermal equilibrium\footnote{A third, possibly speculative mechanism for producing highly non-thermal H$_{2}$ populations is excess energy associated with the dissociation of more complex molecules.  Rovibrationally excited H$_{2}$ and CO have recently been observed in the spectra of comets, consistent with these molecules being the photodissociation products of formaldehyde (H$_{2}$CO; Feldman et al. 2009).}:  excess energy associated with H$_{2}$ formation on dust grains (``formation pumping'') and excitation by UV photons at a rate faster than the molecules can decay via rovibrational emission lines or collisions (``multiple pumping'').  
Initial evidence for a non-thermal component to the CTTS H$_{2}$ population was 
observed by~\citet{herczeg02}, who found H$_{2}$ emission lines pumped out of states near the collisional dissociation threshold of H$_{2}$ in STIS spectra of TW Hya.  These states cannot be thermally populated, and a local source of highly excited H$_{2}$ ([$v$,$J$] = [5,18]) is required if these levels persist.  Dust grains are widely believed to be the primary site of H$_{2}$ formation in the present-day Universe~\citep{hollenbach70,cazaux04} and are abundant in protoplanetary disks (see reviews by Natta et al. 2007 and Dullemond \& Monnier 2010).  Numerous authors have presented H$_{2}$ formation rates in molecular clouds (e.g., Jura 1975; Black \& van Dishoeck 1987; Draine \& Bertoldi 1996).  
Protoplanetary disk models that include molecular formation (e.g., Nomura et al. 2007) have considered the effects of grain growth, which can create significant deviation from the properties of typical interstellar dust grains~\citep{vasyunin11}.  However, these authors do not explicitly describe the distribution of newly formed molecules in the inner disk.  
\citet{draine96} find typical formation states at high rovibrational levels $\langle$[$v$,$J$]$\rangle$~$\approx$~[5,9], with $\langle$$E_{form}$$\rangle$~$\approx$~2.9 eV.  This may suggest that the high energy excursions from a thermal distribution we observe in Figure 3 are the beginning of a higher energy population of H$_{2}$ recently formed in the disk itself.   Additional calculations of the spatial distributions of newly formed H$_{2}$ in CTTS disks would be useful.\nocite{jura75,black87,natta07,nomura07,dullemond10,feldman09}    

A calculation of H$_{2}$ photo-excitation in the AA Tau disk is more straightforward, and here we provide a simple example of the importance of this process.  
The relative importance of UV line and continuum emission to the photoexcitation of trace molecules (CN, HCN, H$_{2}$O) has been examined in some detail (e.g., Bergin et al. 2003).\nocite{bergin03}
We argue that UV photo-excitation of H$_{2}$ can be important even in the absence of a strong Ly$\alpha$ radiation field or a population of $T$(H$_{2}$)~$\geq$~2000~K molecular gas.  The far-UV continuum produced in the accretion flow and magnetically active stellar atmosphere provides sufficient pumping flux to photo-excite even a cold molecular gas.   Far-UV continuum spectra can now be routinely measured in CTTS systems with COS, and following the analyses of~\citet{france11a,france11b}, we can fit the 912~--~2000~\AA\ continuum spectrum from AA Tau.  We find that the integrated far-UV continuum radiation field is $G_{AA}$ = 10$^{6.3}$ $G_{o}$, where $G_{o}$ is the local UV interstellar radiation field ($G_{o}$~=~1.6~$\times$~10$^{-3}$ erg cm$^{-2}$ s$^{-1}$; Habing 1968).\nocite{habing68}  This corresponds to a far-UV continuum luminosity of $L^{cont}_{FUV}$~$\approx$~9~$\times$~10$^{30}$ erg s$^{-1}$.   

Consider an individual H$_{2}$ absorption line that will have a large population fraction even at low-temperatures ($T$(H$_{2}$)~$\leq$~500 K), the Lyman band (4~--~0)R(1) absorption line at 1049.96~\AA.  The UV photon absorption rate, $C_{\lambda}$ (in units of photons s$^{-1}$), can be calculated by 
\begin{equation}
C_{\lambda}~=~ I_{\lambda}~d\lambda~\sigma_{H_{2}}
\end{equation}
where $I_{\lambda}$ is the incident continuum photon flux ($I_{1050}$~$\approx$~2~$\times$~10$^{10}$ photons cm$^{-2}$ s$^{-1}$ \AA$^{-1}$ at 1 AU for AA Tau), $d\lambda$ is the equivalent width of the (4~--~0)R(1) line, and $\sigma_{H_{2}}$ is the line-center absorption cross-section of an individual molecule, 
given by 
\begin{equation}
\sigma_{H2}~=~ \frac{\sqrt{\pi}~e^{2}}{m_{e}~c~b_{H2}}~\lambda_{i}~f_{i}
\end{equation}
\citep{mccandliss03,cartwright70} where $\lambda_{i}$~=~1049.96~\AA\ and $f_{i}$ is the oscillator strength ($\approx$~0.0155).  $\sigma_{H2}$ of the (4~--~0)R(1) transition is 1.2~$\times$~10$^{-14}$ cm$^{-2}$, which is typical of strong H$_{2}$ absorption lines at $\lambda$~$<$ 1100~\AA\ for $b_{H2}$~=~2 km s$^{-1}$.  $d\lambda$ $>$ 0.1~\AA\ for all temperatures when $N$(H$_{2}$)~$\gtrsim$~10$^{18}$ cm$^{-2}$.  

We can incorporate these values to further simplify Eqn 1 for the case of AA Tau: 
\begin{equation}
C_{1050}~=~ (2.4~\times~10^{-5})~\left( \frac{1 AU}{r} \right)^{2}~\left(\frac{\sigma_{H_{2}}}{10^{-14} cm^{2}}\right)
\end{equation}
where $r$ is the radial distance (in AU) of the H$_{2}$ from the star.  To first order, when $C_{\lambda}$ is larger than the transition probabilities ($A$) of the rovibrational H$_{2}$ emission lines, one expects multiple UV photo-excitations before the gas can relax via line emission.  The average transition probability for the quadrupolar H$_{2}$ emission lines in the 1~--~4~$\mu$m bandpass $\langle$$A^{H_{2}}_{IR}$$\rangle$~$\sim$~5~$\times$~10$^{-7}$ s$^{-1}$.  The pure rotational mid-IR emission lines are another 1~--~3 orders of magnitude slower.  Therefore, we see that $\geq$ 40 far-UV continuum photons are absorbed for every one IR emission, in the absence of collisional de-excitation.  

While it is clear that UV photons are important in the level populations of disk H$_{2}$, it is not clear that they are the dominant factor in this balance. 
The critical density is the ratio of the radiative lifetime of a given state (the Einstein $A$ value, with units of s$^{-1}$) and the collision rate for de-excitation out of that state, $\Gamma$ (in units of cm$^{3}$ s$^{-1}$).
The critical densities for most rovibrational states of H$_{2}$ are $<$~10$^{4}$ cm$^{-3}$ at $T$~$\geq$~2000 K~\citep{mandy93}.  H$_{2}$ + H collision rates ($\Gamma_{H2}$) for gas in LTE in the range 2000~--~4500~K are on the order of 10$^{-12}$~--~10$^{-11}$ cm$^{3}$ s$^{-1}$ for the (1~--~0)S($J^{''}$) near-IR rovibrational transitions that can be observed from the ground~\citep{mandy93}.  Therefore at an average density of $\langle$$n_{H_{2}}$$\rangle$~$\approx$~10$^{5}$ cm$^{-3}$ (\S4.2), the collisional de-excitation transition rate is $\Gamma_{H2}$$\langle$$n_{H_{2}}$$\rangle$~$\approx$~10$^{-7}$~--~10$^{-6}$ s$^{-1}$ for this gas, somewhat less than our estimates of the photoexcitation rate by continuum photons at 1 AU.  In a high-density region such as the inner disk atmosphere of a CTTS, strong UV pumping will compete with collisional excitation and de-excitation to produce level populations that appear neither exactly thermal nor purely fluorescent~\citep{draine96}.   

The observations we present in Figure 3 agree with the qualitative predictions of the irradiated disk models of~\citet{nomura05} and~\citet{nomura07}. In these models, the UV radiation field thermalizes the level populations up to a certain $T_{exc}$ through strong photo-electric heating.  Beyond this thermalization energy, the level populations are determined by the UV pumping rates for the relevant transitions.  In our COS observations of AA Tau, we observe a higher thermalization threshold ($T_{exc}$~$>$~2~$\times$~10$^{4}$ K vs. $T_{exc}$~$\lesssim$~1~$\times$~10$^{4}$ K) than predicted by Nomura et al. (2007; Figs. 12 and 13); however, this can most likely be attributed to the radial distances where these distributions are evaluated.  The model calculations are evaluated at 30~--~50~AU, while the H$_{2}$ absorption we observe in AA Tau may be occurring in the inner disk (\S4.2).  At smaller radial distances, the more intense UV radiation field will enhance the photo-electric heating rate, thermalizing the H$_{2}$ population to higher $T_{exc}$.    


\begin{figure}
\begin{center}
\epsfig{figure=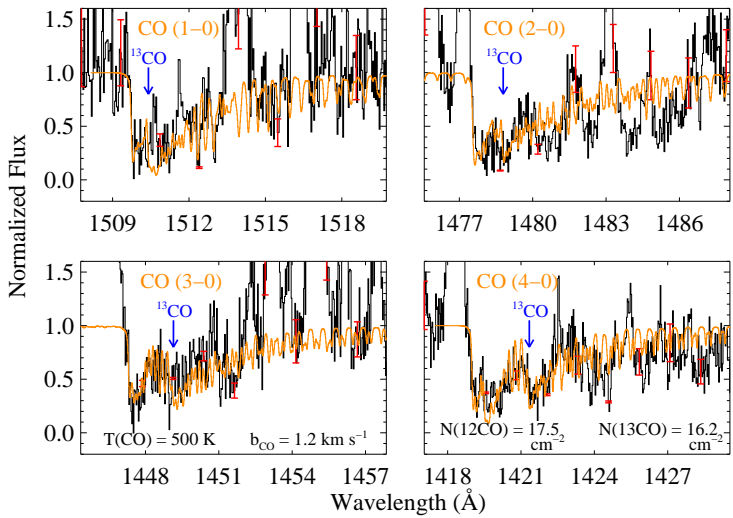,width=2.65in,angle=90}
\vspace{+0.05in}
\caption{Select CO $A$~--~$X$ ($v^{'}$~--~0) absorption bands in the AA Tau spectra.  The data are plotted in black and representative error bars are displayed in red.  These bands were fit simultaneously and the best-fit model (\S3.3) is shown overplotted in orange.  The model parameters are log$_{10}$($N$($^{12}$CO))~=~17.5 cm$^{-2}$, log$_{10}$($N$($^{13}$CO))~=~16.2 cm$^{-2}$, $T$(CO)~=~500 K, and $b_{CO}$~=~1.2 km s$^{-1}$.  
The high temperature required to fit the data argues strongly for an inner disk origin for the absorbing CO gas. 
 }
\end{center}
\end{figure}

\subsection{The Spatial Distribution of the Absorbing Gas}

We have measured the radial velocities of the absorption and emission components of both molecules.  We find that the H$_{2}$ absorption lines are shifted by a uniform +29~$\pm$~8 km s$^{-1}$, slightly higher than the photospheric radial velocity measurement of~\citet{bouvier07} (+17 km s$^{-1}$) and consistent with the atmospheric Balmer-series absorption velocity of +34 km s$^{-1}$ from~\citet{joy49}.  A sample of 10 H$_{2}$ emission lines located both shortward and longward of the Ly$\alpha$ emission line show an H$_{2}$ emission radial velocity of +18~$\pm$~8 km s$^{-1}$, consistent with both the narrow optical line velocity (+4~--~12 km s$^{-1}$) of~\citet{bouvier07} and the H$\alpha$ emission velocity (+13 km s$^{-1}$) from~\citet{joy49}.  Due to the spectral overlap of the CO lines, the radial velocities of this species are not as well constrained.  Furthermore, emission lines coincident with the CO absorption bandheads complicate this sharp edge in the data.  The average CO absorption radial velocity of the $v^{'}$~=~1, 2, 3, 4, and 7 bands is~+20 $\pm$~13 km s$^{-1}$.  The CO emission radial velocity, determined from the bandhead of the $A$~--~$X$ (14~--~4)  complex is +22 $\pm$ 19 km s$^{-1}$.   Therefore, we conclude that there is no evidence for high-velocity infall or outflow in the molecular material we observe towards AA Tau.  

The thermalized H$_{2}$ and CO populations have significantly different excitation temperatures; however, if these absorption lines are both formed in the disk, then they are most likely in close spatial proximity, if not actually co-spatial.  We can put a rough limit on the inner radius of the absorbing gas using the line-widths of the fluorescent H$_{2}$ emission.  We fit the 12 strongest H$_{2}$ emission lines in the 1260~--~1300~\AA\ bandpass, finding $\langle$FWHM$\rangle$~=~53~$\pm$~12 km s$^{-1}$, which corresponds to $r_{in}$~$\approx$~0.16 AU (assuming Keplerian rotation and ignoring thermal and turbulent contributions to the line-width) for a 0.53~M$_{\odot}$ central star~\citep{gullbring98} and a disk inclination of 75\arcdeg~\citep{andrews07}.  Taking the half-width at zero intensity (HWZI) of the line moves $r_{in}$ to $\approx$~0.11 AU. 
 An approximate outer radius of the absorbing material could be where the lower bound on the CO temperature (300 K) is in equilibrium with the radiation field, $r_{out}$~$\approx$~0.79 AU, assuming $T_{*}$~=~4000 K and $R_{*}$~=~1.9 R$_{\odot}$ (from the AA Tau SED fitting of~O'Sullivan et al. 2005).\nocite{osullivan05}  While these inner and outer radii probably do not exactly represent the regions subtended by the absorbing molecules,  they provide an order-of-magnitude estimate of the molecular region.  

The average values of the CO/H$_{2}$ ratio and H$_{2}$ density in the region probed by our absorption line spectroscopy are then: $\langle$CO/H$_{2}$$\rangle$~$\equiv$~$N$(CO)/$N$(H$_{2}$)~$\sim$~0.4 [$^{+2.1}_{-0.3}$] and $\langle$$n_{H_{2}}$$\rangle$ $\equiv$ $N$(H$_{2}$)/($r_{out}$ - $r_{in}$)~$\approx$~8.4~$\times$~10$^{4}$ cm$^{-3}$.   Assuming an H$_{2}$ ortho-para ratio of 3,  the total ortho- + para-H$_{2}$ collision rate, summed over all possible lower levels, is $\Gamma_{TOT}$~=~2.85~$\times$~10$^{-10}$ cm$^{3}$ s$^{-1}$ for $J$~=~9~\citep{yang10}.  The CO critical density for $J$~$\leq$~9 at 300 K is 
 $n_{H_{2}}$~$\approx$~1.8~$\times$~10$^{5}$ cm$^{-3}$.  
Therefore, it is possible that the high-$J$ CO populations that drive the determination of $T_{rot}$(CO) are significantly sub-thermal. 
While the high-J levels are populated by collisions, these levels can radiatively depopulate fast enough that the populations are not representative of the local kinetic temperature. 
To explore this possibility, we computed RADEX models~\citep{vandertak07} using the on-line interface\footnote{ {\tt http://www.sron.rug.nl/$\sim$vdtak/radex/radex.php} }.  
For the values of $T$(H$_{2}$), $N$(CO), $b_{CO}$, and $\langle$$n_{H_{2}}$$\rangle$ found above, we find that CO levels $J$~$\geq$~9 are characterized by temperatures less than the 2500 K kinetic temperature.  For 9~$\leq$~$J$~$\leq$~38, the CO excitation temperature is 483~$\pm$~129 K, in agreement with the $T_{rot}$(CO)~$\sim$~500 K derived in \S3.3.  Therefore, the CO and H$_{2}$ we observe in absorption are consistent with an origin in the same 2500 K gas.   

The average H$_{2}$ number density derived above assumes crude limits to the spatial distribution of the absorbing H$_{2}$ and CO, it is instructive to also consider the limiting cases. At the high-density limit, one might imagine that the entire column of absorbing H$_{2}$ is in the form of a proto gas-giant planet.  Taking ($r_{out}$ - $r_{in}$)~=~1 $d_{*}$, we see that $n_{H_{2}}$~$\sim$~3~$\times$~10$^{6}$ cm$^{-3}$.  At the low-density limit, we might imagine that the H$_{2}$ uniformly spans the entire planet-forming region of the AA Tau disk, or 
($r_{out}$ - $r_{in}$)~=~10 AU.  In this limit, we find $n_{H_{2}}$~$\sim$~5~$\times$~10$^{3}$ cm$^{-3}$ and estimate a total H$_{2}$ mass\footnote{ Assuming $z$~=~0.1~$r$, the volume of this disk is then [(4$\pi$/30)~$\times$~($r_{out}^{3}$~-~$r_{in}^{3}$)].} of $M_{H2}$~$\sim$~2.4~$\times$~10$^{22}$ g, which is roughly 4~$\times$~10$^{-6}$ $M_{\oplus}$.  This calculation confirms our earlier speculation that these line-of-sight measurements do not probe the bulk of the molecular distribution available for planet formation, and are only sampling the tenuous upper regions of the protoplanetary disk.     

\subsubsection{Relation to CO Fundamental Emission}

As noted in \S2.1, CO emission from AA Tau has been detected in fundamental band emission near 5~$\mu$m (Carr \& Najita 2008; Najita et al. 2009; and see also Salyk et al. 2011).\nocite{salyk11}  Our UV absorption-derived CO column densities and temperatures are roughly consistent with those derived from the mid-IR emission lines log$_{10}$($N$(CO))~=~17.7$^{+0.1}_{-0.2}$ and $T$(CO)~=~900~$\pm$~100~K~\citep{carr08}.  This suggests that both observations may be tracing a similar molecular population.  \citet{najita09} find CO line widths of FWHM~=~145 km s$^{-1}$, suggesting that the emission arises from gas in Keplerian rotation within 0.5 AU of the central star (and likely with a significant contribution from gas within 0.1 AU).  Due to the pencil-beam nature of our absorption line measurements, we do not have information regarding the velocity broadening of the CO.  The CO emission line widths are $\sim$~3 times those found for the UV emission lines of H$_{2}$ described in the previous subsection, meaning that the CO emission is likely originating interior to the H$_{2}$ in AA Tau.  $N$(CO) in the range 10$^{17-18}$ cm$^{-2}$ should be associated with H$_{2}$ column densities $N$(H$_{2}$)~$>$~10$^{21}$ cm$^{-2}$, several orders of magnitude larger than we observe.  This discrepancy  raises questions about both the physical origin of the molecular gas and its composition, and these points are addressed in the following subsection.  


\subsection{CO/H$_{2}$ in the AA Tau Circumstellar Disk}

We find that the average observed CO/H$_{2}$ ratio in the AA Tau inner disk is~$\sim$~0.4$^{+2.1}_{-0.3}$.  We imagine two scenarios that could account for this result.  In the first, the absorbing CO resides in a relatively low-density medium ($n_{H2}$~$\lesssim$~2~$\times$~10$^{5}$ cm$^{-3}$), where the higher rotational states that drive the CO temperature fits are sub-thermal.  In this case, the kinetic temperature of the gas is $\sim$~2500 K, the CO and H$_{2}$ are cospatial, and CO/H$_{2}$~$\sim$~0.4.  
A CO/H$_{2}$ ratio in the range~0.1~--~1 is significantly larger than the typical interstellar dense cloud value ($\sim$~10$^{-4}$; Lacy et al. 1994 and see Liszt et al. 2010 for a review of CO/H$_{2}$ in various phases of the molecular ISM).\nocite{lacy94,liszt10}  The collapse of a dense cloud precedes the formation of the disk and protostar, thus finding CO/H$_{2}$ $\geq$ three orders of magnitude above this value suggests significant chemical evolution has taken place in the AA Tau disk.  CO/H$_{2}$ values of $\sim$~unity were recently suggested by~\citet{france11b} in CTTS inner disk atmospheres, using an orthogonal analysis comparing CO emission to typical $N$(H$_{2}$) values from the literature.

In the second scenario, the H$_{2}$ and CO abundance ratios are interstellar, but the absorbing H$_{2}$ is spatially separate from the absorbing CO.  Any CO in the ``hot H$_{2}$ region'' falls below our detection limit and any H$_{2}$ in the ``warm CO region'' is too cool to create measurable opacity at Ly$\alpha$.   
An implication of this reasoning is that there is a reservoir of warm H$_{2}$ associated with the CO absorption, with typical dense cloud abundances. 
Taking the best-fit CO column density (log$_{10}$($N$(CO)) = 17.5 cm$^{-2}$), and assuming a molecular fraction ($f_{H2}$) of 0.67 and CO/H$_{2}$~=~10$^{-4}$, we expect a neutral hydrogen column density of log$_{10}$($N$(H)) = 21.5 cm$^{-2}$ co-spatial with the absorbing CO gas.   The neutral hydrogen opacity associated with 10$^{21.5}$ cm$^{-2}$ would completely extinguish the flux from the star to $>$ $\pm$~5~\AA\ from the Ly$\alpha$ line center, which is inconsistent with the Ly$\alpha$ profile presented in Figure 2a.   
While $f_{H2}$~=~0.67 is ruled out, this scenario is tenable for $f_{H2}$ $\gtrsim$~0.9.  It seems plausible that the molecular fraction is high in a warm ($\sim$~500K) medium with $N$(H$_{2}$)~$>$~10$^{21}$ cm$^{-2}$.  \citet{burgh10} present a direct comparison between the CO/H$_{2}$ ratio and $f_{H2}$ across a range of diffuse and translucent interstellar sightlines.  Extrapolating their Figure 2, one sees that the CO/H$_{2}$~$\approx$~10$^{-4}$ is reached at $f_{H2}$~$\approx$~0.9.  While it would be speculative to draw larger conclusions based on analogy to cold interstellar clouds, it seems that a spatially stratified medium is consistent with our observations.  

Is this vast reservoir of warm H$_{2}$ observable?  At $T$(H$_{2}$) $\lesssim$~500 K, the upper rovibrational states of the molecule are not appreciably populated.  There is not enough opacity in the traditional $HST$ far-UV bandpass ($\lambda$~$\geq$ 1150 \AA) to create a detectable signal.  At $\lambda$~$<$~1120~\AA\ the H$_{2}$ opacity should rise rapidly due to large populations in the low-lying rovibrational levels of the ground electronic state, producing an observable absorption signature that could directly distinguish between the two scenarios laid out above.  Additionally, at $\lambda$~$<$~1120~\AA\ far fewer H$_{2}$ emission lines pumped by Ly$\alpha$ are present~\citep{wood04}, and only one H$_{2}$ line pumped by \ion{O}{6} contributes significantly\footnote{The $C$~--~$X$ (1~--~3) $Q$(3) $\lambda$1119.08 line may be strong, however the branching ratio of the $C$~--~$X$ (1~--~2) $Q$(3) $\lambda$1075.03 line is small ($<$~0.004)}.  The decrease in spectral confusion may offset the decrease in source signal and instrumental effective area across this bandpass. 
With the installation of COS, $HST$ can now observe at wavelengths as short as the Lyman limit~\citep{mccandliss10}, and new medium resolution modes (such as the G130M $\lambda$1222) are in development~\citep{osterman10}.  Heavily damped line profiles require neither high S/N nor high resolution for a robust column density determination, and we suggest that S/N as low as 2 per pixel may be sufficient when the data is rebinned to one spectral resolution element.  
Therefore, we predict that it may be possible to directly search for warm H$_{2}$ in the AA Tau disk with future far-UV observations.   


\section{Summary}

We have presented new spectroscopic observations of the AA Tau protoplanetary disk system. We directly observe both H$_{2}$ and CO absorption lines along the line-of-sight through a CTTS disk for the first time.  These {\it Hubble Space Telescope} observations allow us to constrain the molecular column densities, relative abundances, and temperatures of the two main molecular constituents of the disk.  We find that the H$_{2}$ absorption seen imposed upon the stellar Ly$\alpha$ profile has both thermal and non-thermal components, with the thermal material 
characterized by log$_{10}$($N$(H$_{2}$)) $\approx$ 17.9 cm$^{-2}$ and a temperature of $T$(H$_{2}$) $\sim$ 2500 K.  
The CO absorption implies log$_{10}$($N$(CO)) $\approx$ 17.5 cm$^{-2}$ and $T_{rot}$(CO) $\sim$~500 K.  Taken 
together, the observed CO/H$_{2}$ ratio of $\sim$~0.4 may either be a direct measure of local abundances or indicate spatially stratified  hot and warm molecular layers.  Additional modeling of UV photo-excitation and in situ molecular formation will be very useful for understanding the physical state of the low-density molecular gas in these systems.   

\acknowledgments
K.F. appreciates the hospitality of Poker Flat Research Range, operated by the University of Alaska's Geophysical Institute, where a portion of this work was carried out.  We thank Naseem Rangwala for helpful discussions about CO level populations, M. Eidelsberg for providing CO wavelengths and oscillator strengths used in this work, and an anonymous referee for suggestions that improved the quality of the paper.  This work makes use of data obtained through $HST$ guest observing program 11616 and was supported by NASA grants NNX08AC146 and NAS5-98043 to the University of Colorado at Boulder.  H.A. and E.R. acknowledge the partial support of the Agence Nationale de La Recherche (ANR, France) through the contract ``SUMOSTA'', N0. 09-BLAN-020901

\appendix  

\section{Absorption Model and Non-detection of H$_{2}$ Against the Far-UV Continuum of AA Tau} 

In this Appendix, we describe the creation of an H$_{2}$ model and use it to place limits on absorption against the far-UV continuum of AA Tau. 
The physical characteristics of the H$_{2}$ population in the disk can be constrained by comparing model H$_{2}$ absorption spectra with the data.  Assuming that the H$_{2}$ level populations are controlled by collisional processes~\citep{herczeg04},  
synthetic absorption line spectra are constructed given values of the total H$_{2}$ column density ($N$(H$_{2}$)), the rovibrational excitation temperature ($T$(H$_{2}$)), and Doppler broadening parameter ($b_{H2}$). 
While in principle H$_{2}$ absorption lines can be searched for from the short-wavelength cut-off of our observations ($\lambda$~$\sim$~1133~\AA) to the end of the significant rovibrational population ($\approx$~1400~\AA\ for $T$(H$_{2}$)~$<$~3000 K), practical considerations make the 1225~--~1300~\AA\ band preferred for constraining H$_{2}$ absorption against the continuum of an accreting pre-main sequence star.  At $\lambda$~$<$~1200~\AA, circumstellar and interstellar reddening are the most severe, the continuum used as a background source is declining~\citep{france11a}, and the reflectivity of the $HST$ Optical Telescope Assembly (OTA) is falling precipitously.  All of these factors conspire to create low S/N at these wavelengths.  At $\lambda$~$>$~1300~\AA, the rovibrational levels [$v$,$J$] are not significantly populated, thus only weak absorptions are predicted.   

The 1225~--~1300~\AA\ band is at the peak of the COS effective area and therefore 
provides the best combination of sensitivity and rovibrational population in which to search for an H$_{2}$ absorption signal.  The primary limiting factors for this work are the S/N of the continuum and spectral overlap with the numerous emission features from hot gas lines (e.g., \ion{N}{5} $\lambda$ 1239, 1243) and H$_{2}$ photo-excited by Ly$\alpha$ and \ion{O}{6}~\citep{wilkinson02,herczeg06,france11a}.   We created absorption spectra in the 900~--~1400~\AA\ wavelength range using the H$_{2}$ools optical depth templates~\citep{mccandliss03} for the identification and analysis of H$_{2}$ lines in the COS bandpass.   

H$_{2}$ absorption is not detected against the far-UV continuum of AA Tau.  In order to constrain the column density of the H$_{2}$, we define an upper limit as the column density which produces absorption line depths greater than the 1~$\sigma$ error bars on the continuum. In the 1225~--~1300~\AA\ wavelength range, $N$(H$_{2}$) $<$ 1~$\times$~10$^{19}$ cm$^{-2}$ for $T$(H$_{2}$)~$\geq$~2000 K.  At $T$(H$_{2}$) $<$ 2000 K, we do not expect a significant number of detectable absorbers redward of Ly$\alpha$, therefore this analysis is not sensitive to a high-column, low-temperature H$_{2}$ component in the outer disk (see \S4.3).
At $T$(H$_{2}$)~$>$~2000 K, additional absorbers are predicted that are not observed, and the increased population in the high [$v$,$J$] states lowers the upper limit on $N$(H$_{2}$).
The 2000 K limit also implies $b_{H2}$~$\geq$~4 km s$^{-1}$ because this is the thermal width in the absence of turbulent velocity broadening.  In fact, if turbulence is significant, then the column density limits become smaller because increasing $b_{H2}$ increases the absorption depth at line-center for a given $N$(H$_{2}$).  This 
result is due to the fact that while these lines do not appear to reach zero (``become black'') at their line centers when convolved with the COS LSF, the strongest lines are indeed saturated.  As $b_{H2}$ increases, the equivalent width of the saturated line increases, producing less line-center transmission for a given column density.  We see then that our limit log$_{10}$($N$(H$_{2}$))~$<$~19.0 cm$^{-2}$ is rather conservative because higher temperatures and turbulent velocities only act to reduce this value.

\bibliography{ms_emapj_aatau}



\end{document}